\documentstyle[12pt,amssymb,epsf,epsfig]{article}
\makeatletter
\def\ga{\mathrel{\raise.3ex\hbox{$>$\kern-.75em\lower1ex\hbox{$\sim$}}}}
\def\la{\mathrel{\raise.3ex\hbox{$<$\kern-.75em\lower1ex\hbox{$\sim$}}}}
\def\beq{\begin{equation}}
\def\eeq{\end{equation}}
\def\bea{\begin{eqnarray}}
\def\eea{\end{eqnarray}}

\begin{document}
\vskip .5cm 
\begin{center}
{\Large \bf
Interpretations of
the NuTeV  $\sin^2 \theta_W$ }
\end{center}
\vskip .8cm

\begin{center}
{\bf  S Davidson   }
\end{center}

\begin{center}{\it Dept. of Physics, University of Durham, DH1 3LE, United Kingdom}
\vskip 20pt
\end{center}
\begin{center}
{\bf Abstract}
\end{center}
\begin{quotation}
  {\noindent\small 
We summarize theoretical explanations of the
three $\sigma$ discrepancy between $\sin^2 \theta_W$ measured
by NuTeV and predicted by the Standard Model global
fit. Possible new physics explanations
({\it e.g.} an unmized $Z'$) are not compelling. 
The discrepancy would be reduced by a positive momentum
asymmetry  $s^-$ in the strange sea; present experimental
estimates of $s^-$ are unreliable or  incomplete.
 Upgrading the NuTeV analysis
to NLO would alleviate concerns that the discrepancy
is a QCD effect.}
\end{quotation}





\section{Introduction}

The NuTeV collaboration studied 
$\nu_\mu$ Deep Inelastic Scattering ($\nu DIS$), and  measured 
$\sin^2 \theta_W$ on-shell, or $m_W^2/m_Z^2$, 
to be  $\sin^2 \theta_W  = 0.2276 \pm 0.0013({\rm stat})
\pm 0.0006({\rm syst})\pm 0.0006({\rm theo})$  ~\cite{NuTeV}.
 This is $\sim 3 \sigma$ from the
world average  $\sin^2 \theta_W  = 0.2226 \pm 0.0004$.
Is this  the long-awaited harbinger of
New Physics? 
 Neutrino $DIS$ is  a notoriously difficult environment in which to
do precision physics---is the discrepancy an overlooked
Standard Model (SM) effect?

Various explanations for this discrepancy have been put forward
\cite{Roy,sspna,Babu,nucl}.  In ref.
\cite{sspna}, we considered electroweak corrections, QCD effects,
new physics in loops and new physics at tree level.

\section{New Physics?}

It is difficult to saturate the NuTeV discrepancy with new physics in
loops; an $O(1 \%)$ effect is needed at NuTeV where $Q^2 \sim 20$ GeV$^2$,
but the new physics must not disrupt the part-per-mil
agreement between the SM and precision tests.  
 We found in
\cite{sspna} that oblique corrections, motivated
versions of the MSSM and modified $Z$ couplings 
\footnote{Some authors~\cite{Babu,Tacheuchi} have reconsidered models
where neutrinos mix with heavy singlets,
thereby reducing their couplings with the $Z$ and $W$ bosons
by a factor $1-\epsilon$ and $1-\epsilon/2$ respectively.
However, 
$\epsilon > 0$  reduces the NuTeV
anomaly  at the price of worsening the global fit \cite{sspna}. 
(Our equations differ from those of \cite{Babu} because
we place $\epsilon$ in different electroweak parameters.)}
cannot separately explain the whole 
NuTeV-LEP discrepancy. It has been observed in
\cite{Tacheuchi} that 
oblique corrections induced by new physics,  {\it and} 
 modified $Z$ couplings,
can fit all the data
\footnote{Bernstein, in these proceedings, has
a different interpretation of \cite{sspna} or \cite{Tacheuchi}.} .

New tree-level physics offers more promising
explanations. A $\sim 1\%$ decrease with respect to the SM
of the coefficient of the operator $(\bar{\nu}_\mu \gamma^\alpha \nu_\mu)
(\bar{q_L}  \gamma_\alpha  q_L)$ is required, and could be provided
by a new $Z'$ boson, or by SU(2) triplet leptoquarks with judiciously
chosen unequal masses. A new $Z'$ must have
negligeable mixing with the $Z$ to satisfy
the oblique parameter and  precision bounds on the Z coupling: \cite{Z'}.
However, a $Z'$ coupled to $e.g.$ $B- 3 L_\mu$  would provide
the required four fermion operator at tree level.
(It would also induce the operator $(\bar{\nu}_\mu \gamma^\alpha \nu_\mu)
(\bar{q_R}  \gamma_\alpha  q_R)$; this is
acceptable  because the coefficient \footnote{ 
This coefficient has the wrong sign in the plots of
\cite{sspna}; the $Z'$ has vector couplings so makes
a negative contribution to both $g_L^2$ and $g_R^2$.
We thank Birgit Eberle for bringing this  to our
attention. }
 of
this operator is measured less accurately by NuTeV. )
 The $Z'$  could have
$m_{Z'} > 600 $ GeV  for $g' \sim 1$, or  if the coupling
is small $g' \sim 10^{-3}$, it could have 2 GeV $< m_{Z'} < $ 10 GeV
consistently with all experimental constraints. A
$Z'$ with $m \simeq 3 $GeV  could  fit the  current
$g-2$ discrepancy \cite{g2}.

\section{Back to the Standard Model}

The NuTeV experiment measures the ratio of ``short''
(= muonless) to ``long'' (with a $\mu$) events for incident
$\nu_\mu$ and $\bar{\nu}_\mu$ beams.  From this
they extract the ratios $R^\nu$ and $R^{\bar{\nu}}$, where
$R^{\nu} = 
{ \sigma (\nu N \rightarrow \nu X)}/
{ \sigma (\nu N \rightarrow \mu X)}$ .
$R^\nu$ is more sensitive than  $R^{\bar{\nu}}$ to $\sin^2 \theta_W$,
 so $\sin^2 \theta_W$ is determined mainly from
$R^\nu$, after an effective ``charm mass'' is extracted
 from $R^{\bar{\nu}}$.
NuTeV uses leading order (LO) parton distribution
functions (pdfs), which are fit to their
data, they assume isospin symmetry ($u^p(x) = d^n(x)$),
and that $q(x) = \bar{q}(x)$ for second generation
quarks. 

A theoretically cleaner ratio, where we studied  the effects
of isospin violation and $s \neq \bar{s}$ is the Paschos-Wolfenstein
ratio (related to $R^\nu$ and $R^{\bar{\nu}}$) :
\begin{eqnarray}
R_{PW} & \equiv & \frac{ \sigma (\nu N \rightarrow \nu X) - 
  \sigma (\bar{\nu} N \rightarrow \bar{\nu} X)}
{ \sigma (\nu N \rightarrow \mu X) - 
  \sigma (\bar{\nu} N \rightarrow \bar{\mu} X)} \nonumber \\
& =&   \frac{1}{2} - \sin^2 \theta_W + \left[ (1.3 + O(\alpha_s))(u^- - d^- - s^-)
\right] ~~~,
\label{RPW}
\end{eqnarray}
where the 1.3 is a simplification (see \cite{sspna}).
The square brackets contain the corrections
that arise if isospin is violated, or if
there is a momentum asymmetry in the strange sea:
$s^- \neq 0$, where $ s^- = \int dx x (s(x) - \bar{s}(x))$.

Most pdf fits assume $s^- = 0$.
This was not imposed in  
ref. \cite{BPZ} (BPZ), who   performed a NLO fit  to all
the cross section data available (this did not include
CCFR and NuTeV). They found that  $s^- \simeq .002$
was  a significantly better  fit ($\Delta \chi^2 = 25$ for 2 additional 
d.o.f.)  than $s^- = 0$.  
 Naively substituting
this into eqn. \ref{RPW}, one finds that
 $\sin^2 \theta_W |_{NuTeV} - \sin^2 \theta_W |_{LEP}$ decreases
to less than two $\sigma$. Realistically,
the effect of $s^-$ on  $\sin^2 \theta_W$  will be reduced by 
experimental cuts and sensitivities.
  NuTeV has published  a LO $ s \neq \bar{s}$ 
fit to their dimuon data \cite{Goncharov}, and found $s^-$
{\it negative}.  It is unclear whether the NuTeV dimuon data
is consistent with the CDHSW cross-section data, which
inconjunction with BCDMS,   drives
$s^-$ positive in the BPZ fit; a refit to all
the data  would be required to determine this. However, the
NuTeV analysis  \cite{Goncharov} had various peculiar features,
as outlined in the (post-publication) note added to \cite{sspna}.
After the appearance of \cite{sspna}, NuTeV
pointed out \cite{new} that according to their analysis \cite{Goncharov},
the asymmetry  had the wrong sign to reduce the  $\sin^2 \theta_W$
discrepancy: $s^- \sim -.0027 \pm .0013$.
 They have recently redone their $s \neq \bar{s}$
analysis at NLO\cite{ICHEP}, and   find a {\it positive}
asymmetry  $s^-  \sim .0003$. These determinations are affected by a
   theoretical uncertainty which is not included 
in the quoted error.
A more detailed discussion of the NuTeV $s^-$ extraction
can be found in the ``note added'' to \cite{sspna}.

$R_{PW}$ is theoretically attractive because the parton distributions
cancel out of the ratio at LO, and the NLO corrections are small.  However,
the $R^\nu$, $R^{\bar{\nu}}$ ratios measured  by NuTeV have some dependence
on the pdfs, which is exacerbated 
by  asymmetries between charged-current and neutral-current, 
or between $\nu$ and $\bar \nu$ events.
Such asymmetries could be induced by
experimental cuts and   by different
 $\nu$, $\bar{\nu}$ spectra.
It is therefore difficult
to estimate the size of the NLO corrections to the $\sin^2 \theta_W$
determination from NuTeV, particularily since NuTeV
fit their LO pdfs to their data, which could absorb
some of the NLO effects. A  NLO analysis of the
NuTeV experiment would be a welcome solution to these concerns.

\end{document}